\documentstyle[aps]{revtex}

\begin{document}
\title{The decays $\bar{B}\rightarrow \bar{K}D$ and $\bar{B}\rightarrow \bar{K}\bar{%
D}$ and final state interactions}
\author{Fayyazuddin}
\address{National Center for Physics, Quaid-i-Azam University, Islamabad 45320,\\
Pakistan.}
\maketitle

\begin{abstract}
The decays $\bar{B}\rightarrow \bar{K}D$ and $\bar{B}\rightarrow \bar{K}\bar{%
D}$ taking into account final state interactions are discussed. These decays
are described by four strong phases $\delta _0,\delta _1,\bar{\delta}_0,\bar{%
\delta}_1$ (subscripts 0 and 1 refers to $I=0$ and $I=1$ final states), one
weak phase $\gamma $ and four real amplitudes. It is argued that strong
interaction dynamics implies $\bar{\delta}_1=0,\delta _0=-\delta _1$.
Rescattering has significant effects on weak amplitudes. Taking into
account, rescattering, we find that direct CP--violating asymmetry in these
decays may lie in the range $\mp 0.023\sin \gamma \leq {\cal A}_{1,2}\leq
\mp 0.086\sin \gamma $.
\end{abstract}

The weak decays $\bar{B}\rightarrow \bar{K}D$ and $\bar{B}\rightarrow \bar{K}%
\bar{D}$ taking into account final state interactions have been studied by
several authors \cite{gronau1,gronau2,xing,gronau3}. These decays are
described by four real amplitudes, four strong phases $\delta _0,\delta _1,%
\bar{\delta}_0,\bar{\delta}_1$ and one weak phase $\gamma $. The effective
Lagrangian which describes these decays are given by 
\begin{mathletters}
\label{01}
\begin{eqnarray}
L_{\text{eff}} &=&\frac{G_F}{\sqrt{2}}V_{cb}V_{us}^{*}\left[ \bar{s}\gamma
^\mu \left( 1-\gamma _5\right) u\right] \left[ \bar{c}\gamma _\mu \left(
1-\gamma _5\right) b\right]  \label{01a} \\
L_{\text{eff}} &=&\frac{G_F}{\sqrt{2}}V_{ub}V_{cs}^{*}\left[ \bar{s}\gamma
^\mu \left( 1-\gamma _5\right) c\right] \left[ \bar{u}\gamma _\mu \left(
1-\gamma _5\right) b\right]  \label{01b}
\end{eqnarray}
Since the effective weak Lagrangians for these decays have $\Delta I=1/2$,
the isospin analysis \cite{gronau3} for $\bar{K}D:$%
\end{mathletters}
\begin{mathletters}
\label{02}
\begin{eqnarray}
A\left( B^{-}\rightarrow K^{-}D^0\right) &=&2f_1e^{i\delta _1}  \label{02a}
\\
A\left( \bar{B}^0\rightarrow K^{-}D^{+}\right) &=&\left[ f_1e^{i\delta
_1}+f_0e^{i\delta _0}\right]  \label{02b} \\
A\left( \bar{B}^0\rightarrow \bar{K}^0D^0\right) &=&\left[ f_1e^{i\delta
_1}-f_0e^{i\delta _0}\right]  \label{02c}
\end{eqnarray}
where $\delta _0$ and $\delta _1$ are the phase shifts for $I=0$ and $I=1$
isospin states. On the other hand for $\bar{K}\bar{D}$ we have 
\end{mathletters}
\begin{mathletters}
\label{03}
\begin{eqnarray}
A\left( \bar{B}^0\rightarrow \bar{K}^0\bar{D}^0\right) &=&2\bar{f}%
_1e^{i\gamma }e^{i\bar{\delta}_1}  \label{03a} \\
A\left( B^{-}\rightarrow K^{-}\bar{D}^0\right) &=&e^{i\gamma }\left[ \bar{f}%
_1e^{i\bar{\delta}_1}+\bar{f}_0e^{i\bar{\delta}_0}\right]  \label{03b} \\
A\left( B^{-}\rightarrow \bar{K}^0D^{-}\right) &=&e^{i\gamma }\left[ -\bar{f}%
_1e^{i\bar{\delta}_1}+\bar{f}_0e^{i\bar{\delta}_0}\right]  \label{03c}
\end{eqnarray}
In addition to above decays, other decays relevant to our analysis are 
\end{mathletters}
\begin{eqnarray}
A\left( \bar{B}^0\rightarrow \pi ^{+}D_s^{-}\right) &=&\bar{f}e^{i\gamma
}e^{i\bar{\delta}}=\sqrt{2}A\left( B^{-}\rightarrow \pi ^0D_s^{-}\right)
\label{04} \\
A\left( B^{-}\rightarrow \eta D_s^{-}\right) &=&\frac 1{\sqrt{6}}e^{i\gamma
}\left[ \bar{f}e^{i\bar{\delta}}-2\left( -\bar{f}_1e^{i\bar{\delta}}+\bar{f}%
_0e^{i\bar{\delta}_0}\right) \right]  \label{05}
\end{eqnarray}
where in writting Eq. (\ref{05}), we have used SU(3) which gives 
\begin{equation}
\sqrt{6}A\left( B^{-}\rightarrow \eta D_s^{-}\right) =A\left( \bar{B}%
^0\rightarrow \pi ^{+}D_s^{-}\right) -2A\left( B^{-}\rightarrow \bar{K}^0%
\bar{D}\right)  \label{06}
\end{equation}

In this paper, we address two questions: Is it possible to get some
constraints on strong phases ? How the weak amplitudes are affected by
rescattering ? In answer to first question, our analysis implies 
\begin{equation}
\delta _0=-\delta _1,\,\,\,\,\,\,\bar{\delta}_1=0.  \label{07}
\end{equation}
We find that rescattering has significant effect on some observables for
these decays.

Let us consider the scattering processes: 
\begin{eqnarray*}
\bar{K}+D &\rightarrow &\bar{K}+D \\
\bar{K}+\bar{D} &\rightarrow &\bar{K}+\bar{D}
\end{eqnarray*}
where $\bar{K}\equiv \left( \bar{K}^0,\,K^{-}\right) ,\,D\equiv \left(
D^{+},\,D^0\right) ,\,\bar{D}\equiv \left( \bar{D}^0,\,D^{-}\right) $. We
note that in $s$ or $u$-channel only the states with quark structure $s\bar{c%
}$ can be exchanged ($s,u$ and $t$ are Mandelstam variables). An important
consequence of this is that $s$-channel is exotic for $\bar{K}D$ scattering
whereas $u$-channel is exotic for $\bar{K}\bar{D}$ scattering. Since $s\bar{c%
}$ state has isospin $I=0$, this exchange in $s$-channel gives isospin
projection operator $-2P_0$ where as $I=0$ exchange in $u$-channel gives
isospin projection operator $P_1-P_0$. In $t$-channel, $I=1\left( \rho
\right) $ exchange gives an isospin factor $P_1-3P_0$ where as $I=0\left(
\omega \right) $ exchange gives an isospin factor $P_1+P_0$. Assuming $\rho
-\omega $ degeneracy, $I=1$ and $I=0$ exchanges give $\left( P_1-3P_0\right)
+\left( P_1+P_0\right) =2\left( P_1-P_0\right) $ isospin projection operator
for $\bar{K}D$ scattering where as for $\bar{K}\bar{D}$ scattering. we get $%
\left( P_1-3P_0\right) -\left( P_1+P_0\right) =-4P_0$. Hence for $\bar{K}D$
and $\bar{K}\bar{D}$, the scattering amplitudes can be written as 
\begin{eqnarray}
M &=&-gA\left( P_1-P_0\right)  \label{08} \\
\bar{M} &=&-g\bar{A}\left( -2P_0\right)  \label{09}
\end{eqnarray}
Eqs. (\ref{08}) and (\ref{09}) imply $\delta _0=-\delta _1,\bar{\delta}_1=0$
viz Eq. (\ref{07}).

It is clear from Eq. (\ref{08}) that scattering matrices for $\bar{K}D$ for $%
Q=-1$ and $Q=0$ states are given by respectively 
\begin{equation}
M_{-0}=-gA,\,\,\,\,\,\,\left( K^{-}D^0\rightarrow K^{-}D^0\right) 
\label{10}
\end{equation}
\begin{equation}
M=
\begin{tabular}{c|cc}
& $00$ & $-+$ \\ \hline
$00$ & $0$ & $-gA$ \\ 
$-+$ & $-gA$ & $0$%
\end{tabular}
\label{11}
\end{equation}
From Eq. (\ref{08}), we get for Q=0 and Q=-1 states of $\bar{K}\bar{D}$, the
scattering matrices 
\begin{eqnarray}
\bar{M}_{00} &=&0,\,\,\,\,\,\,\,\,\left( \bar{K}^0\bar{D}^0\rightarrow \bar{K%
}^0\bar{D}^0\right)   \label{12} \\
\bar{M} &=&
\begin{tabular}{ccc}
& $0-$ & $-0$ \\ 
$0-$ & $g\bar{A}$ & $-g\bar{A}$ \\ 
$-0$ & $-g\bar{A}$ & $g\bar{A}$%
\end{tabular}
\label{13}
\end{eqnarray}
However, we note that for $\bar{K}\bar{D}$ scattering other channels $\pi
^{+}D_s^{-},\pi ^0D_s^{-}$ and $\eta D_s^{-}$ are also open. Hence we must
extend the scattering matrices given in Eqs. (\ref{12}) and (\ref{13}) to
take into account the inelastic channels. Using SU(3), we get the scattering
matrices 
\begin{equation}
\bar{M}\left( Q=0\right) =
\begin{tabular}{c|cc}
& $00$ & $\pi ^{+}s^{-}$ \\ \hline
$00$ & $0$ & $-g{A}$ \\ 
$\pi ^{+}s^{-}$ & $-g{A}$ & $0$%
\end{tabular}
\label{14}
\end{equation}
and 
\begin{equation}
\bar{M}\left( Q=-1\right) =
\begin{tabular}{ccccc}
& $0-$ & $-0$ & $\pi ^0s^{-}$ & $\eta s^{-}$ \\ 
$0-$ & $g\bar{A}$ & $-g\bar{A}$ & $\frac 1{\sqrt{2}}gA$ & $\frac 1{\sqrt{6}}%
g\left( 2\bar{A}-A\right) $ \\ 
$-0$ & $-g\bar{A}$ & $g\bar{A}$ & $-\frac 1{\sqrt{2}}gA$ & $\frac 1{\sqrt{6}}%
g\left( 2\bar{A}-A\right) $ \\ 
$\pi ^0s^{-}$ & $\frac 1{\sqrt{2}}gA$ & $-\frac 1{\sqrt{2}}gA$ & 0 & 0 \\ 
$\eta s^{-}$ & $\frac 1{\sqrt{6}}g\left( 2\bar{A}-A\right) $ & $\frac 1{%
\sqrt{6}}g\left( 2\bar{A}-A\right) $ & 0 & $-\frac 23g\left( A+\bar{A}%
\right) $%
\end{tabular}
\label{15}
\end{equation}
To proceed further we note that in terms of Regge phenomenology, exotic $u$%
-channel implies exchange degeneracy i.e. in $t$-channel $\rho -A_2$ as well
as $\omega -f$ trajectories are exchange degenrate. Taking this into
account, it is convenient to express amplitudes $A$ and $\bar{A}$ in
Veneziano representation \cite{venezeno} 
\begin{eqnarray}
A &=&\left[ \frac{\Gamma \left( 1-\alpha _{D_s^{*}}\left( u\right) \right)
\Gamma \left( 1-\alpha \left( t\right) \right) }{\Gamma \left( 1-\alpha
_{D_s^{*}}\left( u\right) -\alpha \left( t\right) \right) }\right] 
\label{16} \\
\bar{A} &=&\left[ \frac{\Gamma \left( 1-\alpha _{D_s^{*}}\left( s\right)
\right) \Gamma \left( 1-\alpha \left( t\right) \right) }{\Gamma \left(
1-\alpha _{D_s^{*}}\left( s\right) -\alpha \left( t\right) \right) }\right] 
\label{17}
\end{eqnarray}
We will take linear Regge trajectories viz 
\begin{eqnarray}
\alpha  &=&\alpha _0+\alpha ^{\prime }t  \nonumber \\
\alpha _{D_s^{*}}\left( u\right)  &=&\alpha _{D_s^{*}}\left( 0\right)
+\alpha ^{\prime }u  \nonumber \\
\alpha _{D_s^{*}}\left( s\right)  &=&\alpha _{D_s^{*}}\left( 0\right)
+\alpha ^{\prime }s  \label{18}
\end{eqnarray}
We assume universal slope viz 
\begin{equation}
\alpha ^{\prime }\simeq 0.94\text{ GeV}^{-2}\approx \frac 1{s_0}  \label{19}
\end{equation}
and take $\alpha _0=0.46$. In actual numerical evaluation we will put $%
\alpha _0=1/2$ and $s_0=1$ GeV$^2$. Note that $\alpha _{D_s^{*}}\left(
0\right) =1-\alpha ^{\prime }m_{D_s^{*}}^2$ for large $s$, we get from Eqs. (%
\ref{16}) and (\ref{17})$:$%
\begin{eqnarray}
A &\rightarrow &\frac \pi {\sin \pi \alpha \left( t\right) \Gamma \left(
\alpha \left( t\right) \right) }\left( \frac s{s_0}\right) ^{\alpha \left(
t\right) }  \label{20} \\
\bar{A} &\rightarrow &\frac{\pi e^{-i\pi \alpha \left( t\right) }}{\sin \pi
\alpha \left( t\right) \Gamma \left( \alpha \left( t\right) \right) }\left( 
\frac s{s_0}\right) ^{\alpha \left( t\right) }  \label{21}
\end{eqnarray}
From Eqs. (\ref{17}), we note that $1-\alpha _{D_s^{*}}\left( s\right) =0$
gives a pole at $s=m_{D_s^{*}}^2$ and $1-\alpha \left( t\right) =0$ gives a
pole at $t=m_\rho ^2$. Using this property of Venezeno representation, we
get 
\begin{equation}
-g=2g_{D_s^{*}DK}^2=4g_{\rho K\bar{K}}g_{\rho D\bar{D}}  \label{22}
\end{equation}
Using the usual parameterization for $g_{D_s^{*}DK}^2$, we get 
\begin{equation}
g_{D_s^{*}DK}^2=\gamma _D^2\frac{m_{D_s^{*}}^2}{f_K^2}\simeq 50  \label{23}
\end{equation}
for $\gamma _D=1/2$.

We now come to the second question viz the effect of rescattering on the
weak amplitudes. First we discuss the rescattering for $\bar{K}D$ system.
Two particle unitarity gives \cite{donoghue,falk,caprini} 
\begin{eqnarray}
Disc\,A\left( \bar{B}^0\rightarrow \bar{K}^0D^0\right) &=&\frac 1{32\pi }%
\frac{\left| {\bf p}\right| }s\int d\Omega A\left( \bar{B}^0\rightarrow
K^{-}D^{+}\right) M^{*}\left( K^{-}D^{+}\rightarrow \bar{K}^0D^0\right) 
\nonumber \\
&=&\frac 1{32\pi }\frac 1{\sqrt{s}\left| {\bf p}\right| }\int_{-2\left| {\bf %
p}\right| ^2}^0dt\,A\left( \bar{B}^0\rightarrow K^{-}D^{+}\right)
M^{*}\left( K^{-}D^{+}\rightarrow \bar{K}^0D^0\right)  \label{24}
\end{eqnarray}
Using Eqs. (\ref{11}) and (\ref{20}), we get 
\begin{equation}
Disc\,A\left( \bar{B}^0\rightarrow \bar{K}^0D^0\right) =\frac 1{32\pi }\frac 
1{\sqrt{s}\left| {\bf p}\right| }\frac{\left( -g\right) \pi A\left( \bar{B}%
^0\rightarrow K^{-}D^{+}\right) }{\Gamma \left( \alpha \left( 0\right)
\right) \sin \pi \alpha \left( 0\right) }\int_{-2\left| {\bf p}\right|
^2}^0e^{\alpha \left( t\right) \ln \left( s/s_0\right) }dt  \label{25}
\end{equation}
where we have put \cite{falk}, $\alpha \left( t\right) \approx \alpha \left(
0\right) $ in $\Gamma \left( \alpha \left( t\right) \right) $ and $\sin \pi
\alpha \left( t\right) $. Hence we get, taking $\alpha \left( 0\right)
\simeq 1/2$ and using linear Regge trajectory: 
\begin{equation}
Disc\,A\left( \bar{B}^0\rightarrow \bar{K}^0D^0\right) \approx -\frac g{16%
\sqrt{\pi }}\frac{\left( s/s_0\right) ^{\alpha _0-1}}{\ln \left(
s/s_0\right) }A\left( \bar{B}^0\rightarrow K^{-}D^{+}\right)  \label{26}
\end{equation}
We now use dispersion relation to obtain $A\left( \bar{B}^0\rightarrow
K^{-}D^{+}\right) $ \cite{donoghue,falk}: 
\begin{eqnarray}
A\left( \bar{B}^0\rightarrow \bar{K}^0D^0\right) _{\text{FSI}} &=&-\frac g{16%
\sqrt{\pi }}\frac 1{\ln \left( m_B^2/s_0\right) }A\left( \bar{B}%
^0\rightarrow K^{-}D^{+}\right) \times \frac 1\pi \int_{\left(
m_D+m_K\right) ^2}^\infty \frac{\left( s/s_0\right) ^{\alpha _0-1}}{s-m_B^2}
\nonumber \\
&=&-\frac g{16\sqrt{\pi }}\frac{\sqrt{s_0}}{m_B}\frac 1{\ln \left(
m_B^2/s_0\right) }\frac 1\pi \left[ i\pi +\ln \frac{1+x}{1-x}\right] A\left( 
\bar{B}^0\rightarrow K^{-}D^{+}\right)  \nonumber \\
&\equiv &\epsilon e^{i\theta }A\left( \bar{B}^0\rightarrow K^{-}D^{+}\right)
\label{27}
\end{eqnarray}
where in evaluating the dispersion integral we have replaced \cite{falk} $%
\ln \left( s/s_0\right) $ by $\ln \left( m_B^2/s_0\right) $. In Eq. (\ref{27}%
) $x,\epsilon $ and $\theta $ are given by 
\begin{eqnarray}
x &=&\frac{m_D+m_K}{m_B}\simeq 0.447  \nonumber \\
\epsilon &=&-\frac g{16\sqrt{\pi }}\frac{s_0}{m_B\ln \left( m_B^2/s_0\right) 
}\sqrt{1+\frac 1{\pi ^2}\left( \ln \frac{1+x}{1-x}\right) ^2}  \nonumber \\
&=&-g\left( 2.01\times 10^{-3}\right) =0.20  \nonumber \\
\theta &=&\tan ^{-1}\left[ \frac \pi {\ln \frac{1+x}{1-x}}\right] \approx
73^0  \label{28}
\end{eqnarray}
where we have used $-g\approx 100$ as given in Eqs. (\ref{22}) and (\ref{23})

Similarly, we obtain 
\begin{equation}
A\left( \bar{B}^0\rightarrow K^{-}D^{+}\right) _{\text{FSI}}=\epsilon
e^{i\theta }A\left( \bar{B}^0\rightarrow \bar{K}^0D^0\right)  \label{30}
\end{equation}
For the decays $\bar{B}\rightarrow \bar{K}\bar{D}$, we note that except for
the decays $\bar{B}^0\rightarrow \pi ^{+}D_s^{-}$, $B^{-}\rightarrow \pi
^0D_s^{-}$, $B^{-}\rightarrow \eta D_s^{-}$, all other decays are either
color suppressed or are given by annihilation amplitude. Thus in evaluating
rescattering correction, we retain only the dominent amplitudes. From Eqs. (%
\ref{14}) and (\ref{15}), following the same procedure as above, we get 
\begin{eqnarray}
A\left( \bar{B}^0\rightarrow \bar{K}^0\bar{D}^0\right) _{\text{FSI}}
&=&\epsilon e^{i\theta }A\left( \bar{B}^0\rightarrow \pi ^{+}D_s^{-}\right)
\label{31} \\
A\left( B^{-}\rightarrow \bar{K}^0D^{-}\right) _{\text{FSI,}\pi } &=&\frac 12%
\epsilon e^{i\theta }A\left( \bar{B}^0\rightarrow \pi ^{+}D_s^{-}\right) 
\nonumber \\
&=&-A\left( B^{-}\rightarrow K^{-}\bar{D}^0\right) _{\text{FSI,}\pi }
\label{32} \\
A\left( B^{-}\rightarrow \bar{K}^0D^{-}\right) _{\text{FSI,}\eta } &=&\frac 1%
{\sqrt{6}}\epsilon e^{i\theta }\left[ 1-2\frac{\ln \left( m_B^2/s_0\right) }{%
\sqrt{\pi ^2+\left( \ln \left( m_B^2/s_0\right) \right) ^2}}e^{i\pi
/2}e^{-i\chi }\right] A\left( \bar{B}^0\rightarrow \eta D_s^{-}\right) 
\nonumber \\
&=&A\left( B^{-}\rightarrow K^{-}\bar{D}^0\right) _{\text{FSI}}  \label{33}
\end{eqnarray}
where 
\begin{equation}
\chi =\tan ^{-1}\frac \pi {\ln \left( m_B^2/s_0\right) }\approx \tan
^{-1}\left( 0.94\right) \approx \frac \pi 4  \label{34}
\end{equation}
In order to simplify the calculation, we replace 
\begin{equation}
\frac{\ln \left( m_B^2/s_0\right) }{\sqrt{\left( \ln \left( m_B^2/s_0\right)
\right) ^2+\pi ^2}}=0.727\approx \frac 1{\sqrt{2}}  \label{35}
\end{equation}
Hence, from Eq. (\ref{33}), using Eqs. (\ref{34}) and (\ref{35}), we get 
\begin{eqnarray}
A\left( B^{-}\rightarrow \bar{K}^0D^{-}\right) _{\text{FSI,}\eta } &=&-\frac 
i{\sqrt{6}}\epsilon e^{i\theta }A\left( \bar{B}^0\rightarrow \eta
D_s^{-}\right)  \nonumber \\
&=&A\left( B^{-}\rightarrow K^{-}\bar{D}^0\right) _{\text{FSI,}\eta }
\label{36}
\end{eqnarray}
Now using Eq. (\ref{06}), and neglecting the contribution of $A\left(
B^{-}\rightarrow \bar{K}^0D^{-}\right) $ compared to $A\left( \bar{B}%
^0\rightarrow \pi ^{+}D_s^{-}\right) $, we obtain 
\begin{eqnarray}
A\left( B^{-}\rightarrow \bar{K}^0D^{-}\right) _{\text{FSI,}\pi ^0-\eta } &=&%
\frac 12\epsilon e^{i\theta }\left( 1-\frac i3\right) A\left( \bar{B}%
^0\rightarrow \pi ^{+}D_s^{-}\right)  \nonumber \\
&=&\frac{\sqrt{10}}6\epsilon e^{i\left( \theta -\phi \right) }A\left( \bar{B}%
^0\rightarrow \pi ^{+}D_s^{-}\right)  \label{37} \\
A\left( B^{-}\rightarrow K^{-}\bar{D}^0\right) _{\text{FSI,}\pi ^0-\eta }
&=&-\frac 12\epsilon e^{i\theta }\left( 1+\frac i3\right) A\left( \bar{B}%
^0\rightarrow \pi ^{+}D_s^{-}\right)  \nonumber \\
&=&-\frac{\sqrt{10}}6\epsilon e^{i\left( \theta +\phi \right) }A\left( \bar{B%
}^0\rightarrow \pi ^{+}D_s^{-}\right)  \label{38}
\end{eqnarray}
where 
\begin{eqnarray}
\phi &=&\tan ^{-1}\left( \frac 13\right) \approx 18^0  \nonumber \\
\theta +\phi &=&91^0\approx 90^0  \nonumber \\
\theta -\phi &=&55^0  \label{39}
\end{eqnarray}
Taking into rescattering, Eqs. (\ref{02}) and (\ref{03}) are modified 
\begin{mathletters}
\label{40}
\begin{eqnarray}
A\left( B^{-}\rightarrow K^{-}D^0\right) &=&2f_1\left( 1+\epsilon e^{i\theta
}\right) e^{i\delta _1}  \label{40a} \\
A\left( \bar{B}^0\rightarrow K^{-}D^{+}\right) &=&f_1\left( 1+\epsilon
e^{i\theta }\right) e^{i\delta _1}+f_0\left( 1-\epsilon e^{i\theta }\right)
e^{i\delta _0}  \label{40b} \\
A\left( \bar{B}^0\rightarrow \bar{K}^0D^0\right) &=&f_1\left( 1+\epsilon
e^{i\theta }\right) e^{i\delta _1}-f_0\left( 1-\epsilon e^{i\theta }\right)
e^{i\delta _0}  \label{40c}
\end{eqnarray}
\end{mathletters}
\begin{mathletters}
\label{41}
\begin{eqnarray}
A\left( \bar{B}^0\rightarrow \bar{K}^0\bar{D}^0\right) &=&e^{i\gamma }\left[
2\bar{f}_1+\epsilon \bar{f}e^{i\left( \theta +\bar{\delta}\right) }\right]
\label{41a} \\
A\left( B^{-}\rightarrow K^{-}\bar{D}^0\right) &=&e^{i\gamma }\left[ \left( 
\bar{f}_1+\bar{f}_0e^{i\bar{\delta}_0}\right) -\frac{\sqrt{10}}6\epsilon
e^{i\left( \theta +\phi +\bar{\delta}\right) }\bar{f}\right]  \label{41b} \\
A\left( B^{-}\rightarrow \bar{K}^0D^{-}\right) &=&e^{i\gamma }\left[ \left( -%
\bar{f}_1+\bar{f}_0e^{i\bar{\delta}_0}\right) +\frac{\sqrt{10}}6\epsilon
e^{i\left( \theta -\phi +\bar{\delta}\right) }\bar{f}\right]  \label{41c}
\end{eqnarray}
The observables which are significantly affected by rescattering can be
easily obtained from Eqs. (\ref{40}) and (\ref{41}). From these equations,
we get 
\end{mathletters}
\begin{eqnarray}
R &=&\frac{\Gamma \left( \bar{B}^0\rightarrow \bar{K}^0\bar{D}^0\right) }{%
\Gamma \left( \bar{B}^0\rightarrow \bar{K}^0D^0\right) }  \nonumber \\
&\simeq &\frac{4\bar{f}_1^2\left[ 1+\epsilon \frac{\bar{f}}{\bar{f}_1}%
+\epsilon ^2\frac{\bar{f}^2}{4\bar{f}_1^2}\right] }{f_1^2+f_0^2-2f_1f_0\cos
\left( \delta _1-\delta _0\right) }  \label{42}
\end{eqnarray}
(where in the denominator, we have neglected the terms containing $\epsilon $%
). 
\begin{eqnarray}
R_{1,2} &\equiv &\frac{\Gamma \left( B^{-}\rightarrow K^{-}D_{1,2}\right)
+\Gamma \left( B^{+}\rightarrow K^{+}D_{1,2}\right) }{\Gamma \left(
B^{-}\rightarrow K^{-}D^0\right) }  \nonumber \\
&=&1+\frac 14\left[ r_1^2+r_0^2+2r_1r_0\cos \bar{\delta}_0\right] \mp \cos
\gamma \left[ r_1\cos \delta _1+r_0\cos \left( \delta _1-\bar{\delta}%
_0\right) \right] -\frac{\sqrt{10}}6\epsilon r\cos \left( \theta +\phi +\bar{%
\delta}-\delta _1\right)  \label{43}
\end{eqnarray}
\begin{eqnarray}
{\cal A}_{1,2} &\equiv &\frac{\Gamma \left( B^{-}\rightarrow
K^{-}D_{1,2}\right) -\Gamma \left( B^{+}\rightarrow K^{+}D_{1,2}\right) }{%
\Gamma \left( B^{-}\rightarrow K^{-}D^0\right) }  \nonumber \\
&=&\mp \sin \gamma \left[ r_1\sin \delta _1+r_0\sin \left( \delta _1-\bar{%
\delta}_0\right) +\frac{\sqrt{10}}6\epsilon r\sin \left( \theta +\phi +\bar{%
\delta}-\delta _1\right) \right]  \label{44}
\end{eqnarray}
where $D_{1,2}=\left( D^0\mp \bar{D}^0\right) /\sqrt{2}$ are CP-eigenstates
with CP $=+1,-1$ respectively. $r_1,r_0$ and $r$ are given by 
\begin{equation}
r_1=\frac{\bar{f}_1}{f_1},\text{ }r_0=\frac{\bar{f}_0}{f_1}\text{ and }r=%
\frac{\bar{f}}{f_1}  \label{45}
\end{equation}

So far our analysis is general. To proceed further, we note that \cite
{neubert} that these decays are determined by the tree amplitudes $T\left( 
\bar{T}\right) $, the color suppressed amplitudes $C\left( \bar{C}\right) $
and annihilation amplitude $\bar{A}$. In terms of these amplitudes 
\begin{eqnarray}
f_1 &=&\frac{G_F}{\sqrt{2}}\left| V_{cb}V_{us}^{*}\right| \frac 12\left(
T+C\right)  \label{46} \\
f_0 &=&\frac{G_F}{\sqrt{2}}\left| V_{cb}V_{us}^{*}\right| \frac 12\left(
T-C\right)  \label{47} \\
\bar{f}_1 &=&\frac{G_F}{\sqrt{2}}\left| V_{ub}V_{cs}^{*}\right| \frac 12\bar{%
C}  \label{48} \\
\bar{f}_0 &=&\frac{G_F}{\sqrt{2}}\left| V_{ub}V_{cs}^{*}\right| \frac 12%
\left( \bar{C}+2\bar{A}\right)  \label{49} \\
\bar{f} &=&\frac{G_F}{\sqrt{2}}\left| V_{ub}V_{cs}^{*}\right| \bar{T}
\label{50}
\end{eqnarray}
Note that in the Wolfenstein representation of CKM matrix \cite{wolfenstein}%
: 
\begin{equation}
\frac{\left| V_{ub}V_{cs}^{*}\right| }{\left| V_{cb}V_{us}^{*}\right| }=%
\sqrt{\rho ^2+\eta ^2}  \label{51}
\end{equation}
In the factorization anastz, these amplitudes are given by \cite{neubert} 
\begin{eqnarray}
T &=&a_1f_KF_0^{B-D}\left( m_K^2\right) \left( m_B^2-m_D^2\right)  \nonumber
\\
C &=&a_2f_DF_0^{B-K}\left( m_D^2\right) \left( m_B^2-m_K^2\right) =\bar{C} 
\nonumber \\
\bar{T} &=&a_1f_{D_s}F_0^{B-\pi }\left( m_{D_s}^2\right) \left( m_B^2-m_\pi
^2\right)  \nonumber \\
\bar{A} &=&a_1f_BF_0^{D-K}\left( m_B^2\right) \left( m_D^2-m_K^2\right)
\label{52}
\end{eqnarray}
where $F_0\left( t\right) $ is scalar form factor for $B$ to $P$ transition $%
\left( P=D,K\text{ or }\pi \right) $. For these form factors, we use the
following values 
\begin{equation}
F_0^{B-K}\left( m_D^2\right) \approx F_0^{B-\pi }\left( m_{D_s}^2\right)
\simeq 0.22  \label{53}
\end{equation}
and \cite{neubert} 
\begin{equation}
F_0^{B-D}\left( m_K^2\right) \approx 0.587  \label{54}
\end{equation}
Using \cite{neubert}, $f_D\simeq 200$ MeV, $f_{D_s}\simeq 240$ MeV, $%
f_B\simeq 180$ MeV, $f_K=158$ MeV and $a_2/a_1\simeq 0.26$, we get 
\begin{eqnarray}
r_1 &=&\frac{\bar{f}_1}{f_1}=\sqrt{\rho ^2+\eta ^2}\frac{\bar{C}}{C+T}%
\approx 0.04  \nonumber \\
r_0 &=&\frac{\bar{f}_0}{f_1}=\sqrt{\rho ^2+\eta ^2}\frac{\bar{C}+2\bar{A}}{%
C+T}\approx 0.08  \nonumber \\
r &=&\frac{\bar{f}}{f_1}=\sqrt{\rho ^2+\eta ^2}\frac{\bar{T}}{C+T}\approx
0.41  \label{55}
\end{eqnarray}
where we have 
\begin{equation}
\sqrt{\rho ^2+\eta ^2}=0.36  \label{56}
\end{equation}

We now discuss the consequences of our main results given in Eqs. (\ref{42})
and (\ref{44}). First we note that it follows from Eq. (\ref{14}), that $%
\bar{\delta}$ can be put equal to zero (no elastic scattering for $\pi
D_s^{-}$). Hence, from Eq. (\ref{42}), we get

\begin{equation}
R=R_0\left[ 1+\epsilon \bar{r}\cos \theta +\frac{\epsilon ^2\bar{r}^2}4%
\right]  \label{57}
\end{equation}
where 
\begin{equation}
R_0=\frac{4\bar{f}_1^2}{f_1^2+f_0^2-2f_1f_0\cos \left( \delta _1-\delta
_0\right) }  \label{58}
\end{equation}
is the branching ratio in the absence of rescattering. From Eqs. (\ref{48}),
(\ref{50}) and (\ref{51}), we get 
\begin{equation}
\bar{r}\equiv \frac{\bar{f}}{\bar{f}_1}=\frac{2\bar{T}}{\bar{C}}\simeq 5.00
\label{59}
\end{equation}
Hence we obtain using $\epsilon =0.20$ and $\theta =73^0,$%
\begin{equation}
R=1.54R_0  \label{60}
\end{equation}
We now discuss the consequence of Eq. (\ref{44}) i.e. direct CP-violation in 
$\bar{B}$ decays.Let us first assume that final state interactions are taken
care of by the phases induced by rescattering. Hence we put $\delta _1$ and $%
\delta _1-\bar{\delta}_0$ equal to zero. Then, we get from Eq. (\ref{44}) 
\begin{eqnarray}
{\cal A}_{1,2} &=&\mp \sin \gamma \left[ \frac{\sqrt{10}}6\epsilon r\sin
\left( \theta +\phi \right) \right]  \nonumber \\
&=&\mp \sin \gamma \left[ \frac{\sqrt{10}}6\epsilon r\right]  \nonumber \\
&=&\mp \sin \gamma \left( 0.043\right)  \label{61}
\end{eqnarray}
since $\theta +\phi \approx 90^0.$

In general however 
\begin{equation}
{\cal A}_{1,2}=\mp \sin \gamma \left[ 0.04\sin \delta _1+0.08\sin \left(
\delta _1-\bar{\delta}_0\right) +0.04\sin \left( \frac \pi 2-\delta
_1\right) \right]  \label{62}
\end{equation}

But the structure of Eqs. (\ref{08}), (\ref{09}), (\ref{16}) and (\ref{17})
implies that $\bar{\delta}_0$ has the same sign as $\delta _0$, it is
therefore reasonable to conclude that $\delta _1-\bar{\delta}_0$ has the
same sign as $\delta _1-\delta _0$. Since $\delta _0=-\delta _1$, it follows
that (since $\delta _1$ has positive sign), that ${\cal A}_{1,2}$ should be
atleast $\mp \left( 0.043\right) \sin \gamma $. The phases $\delta _1,\bar{%
\delta}_0$ are expected to be small. As an example, let us take $\delta
_1=13^0,\delta _0=-13^0=\bar{\delta}_0$, then we get 
\begin{equation}
{\cal A}_{1,2}=\mp \sin \gamma \left( 0.086\right)  \label{63}
\end{equation}

The direct CP--violation is an important consequence of the standard model.
But in the absence of final state interactions, this parameter is zero. Our
analysis shows that even if strong phases $\delta $'s are negligible, the
rescattering gives a finite value for ${\cal A}_{1,2}$ which may be
experimentally measureable in future experiments. Even if our estimate of $%
\epsilon $ is off by a factor 2, ${\cal A}_{1,2}$ will still have at least
the value $\mp 0.023\sin \gamma $. Thus unless in an unlikely case that $%
\bar{\delta}_0$ has the same sign as $\delta _1$ and much greater than $%
\delta _1$, One may conclude that direct asymmetry parameter may lie in the
range 
\begin{equation}
\mp 0.023\sin \gamma \leq {\cal A}_{1,2}\leq \mp 0.086\sin \gamma  \label{64}
\end{equation}

\end{document}